\documentstyle[aps,multicol,tighten,epsfig]{revtex}

\newcommand{\aop}{{\hat{a}}}
\newcommand{\acop}{{{\hat{a}}^{\dagger}}}

\newcommand{\Hop}{{\hat{H}}}

\newcommand{\pop}{{\hat p}}

\newcommand{\sigmaop}{{\hat{\sigma}}}
\newcommand{\xop}{{\hat x}}
\newcommand{\yop}{{\hat y}}
\newcommand{\zop}{{\hat z}}

\newcommand{\Avec}{{\bf A}}   
\newcommand{\bvec}{{\bf b}}
\newcommand{\Bvec}{{\bf B}}

\newcommand{\pvec}{{\bf p}}

\newcommand{\omegac}{{\omega_{c}}}

\newcommand{\omegas}{{\omega_{s}}}
\newcommand{\omegaz}{{\omega_{z}}}

\begin{document}
\draft 

\title{Schr\"odinger-Cat Entangled State Reconstruction in the Penning Trap}

\author{Michol Massini, Mauro Fortunato, Stefano Mancini, Paolo 
Tombesi, and David Vitali}
\address {INFM and Dipartimento di Matematica e Fisica, Universit\`a di 
Camerino \\ via Madonna delle Carceri I--62032  Camerino, Italy}
\date{\today}
\maketitle
\begin{abstract}
We present a tomographic method for the reconstruction of the full 
entangled quantum state for the cyclotron and spin degrees of freedom of an 
electron in a Penning trap. Numerical simulations of the reconstruction of
several significant quantum states show that the method turns out to be
quite accurate.
\end{abstract}

\begin{multicols}{2}

\section{Introduction}
\label{intro}

A single electron trapped in a Penning trap~\cite{Brown1} is a 
unique quantum system in that it allows the measurement of fundamental 
physical constants with striking accuracy. Recently, the electron 
cyclotron degree of freedom has been cooled to its ground state, 
where the electron may stay for hours, and quantum jumps between 
adjacent Fock states have been observed~\cite{gabr}. It is therefore evident 
that the determination of the genuine (possibly entangled) quantum state of
the trapped electron is an important issue, with implications in the very 
foundations of physics, and in particular of quantum mechanics.
After the pioneering work of Vogel and Risken~\cite{vori}, several 
methods have been proposed in order to reconstruct the quantum state 
of light and matter~\cite{phwo}, which range from quantum
tomography~\cite{ray} through quantum state endoscopy~\cite{endo},
to Wigner function determination from outcome probabilities~\cite{davi}.
Also, different techniques~\cite{kag} have been proposed which allow
to deal with entangled states.

In fact, entanglement~\cite{schr} has been defined as one of the most
puzzling features of quantum mechanics, and it is at the heart of quantum
information processing. Some fascinating examples of the possibilities offered
by sharing quantum entanglement are quantum teleportation~\cite{qt93,exp},
quantum dense coding~\cite{benwies}, entanglement swapping~\cite{swap},
quantum cryptography~\cite{crypt}, and quantum computation~\cite{qucomp}.
A striking achievement in this field has been the recent entanglement 
of four trapped ions~\cite{win}.

In the present work we propose to reconstruct the full {\em entangled} state 
(combined cyclotron and spin state) of an electron in a Penning trap 
by using a modified version of quantum state tomography. Previous 
proposals~\cite{sman} need the {\it a priori} knowledge of the
spin state and therefore are not able to deal with entangled states.
Our method, on the contrary, has the ability of measuring the full 
(entangled) pure state of the two relevant degrees of freedom of the
electron. In order to reach this scope, our method takes advantage of
the magnetic bottle configuration to perform simultaneous measurements
of the cyclotron excitation and of the $z$ component of the spin as a
function of the phase of an applied driving electromagnetic field.
The complete structure of the cyclotron-spin quantum state is then
obtained with the help of a tomographic reconstruction from the measured
data.

The present paper is organized as follows: In Sec.~\ref{model} we 
outline the basic model of an electron trapped in a Penning trap, 
while in Sec.~\ref{measure} we describe the main idea of our 
reconstruction procedure. In Sec.~\ref{spin} and \ref{tomo} we 
concentrate on the measurement of the spin and on the tomographic 
reconstruction of the cyclotron states, respectively. We present the 
results of our numerical simulations in Sec.~\ref{simu}, and 
conclude briefly in Sec.~\ref{conclu}.

\section{The basic model}
\label{model}

Let us consider the motion of an electron trapped by the
combination of a homogeneous magnetic field $\Bvec_{0}$ along the
positive $z$ axis and an electrostatic quadrupole potential in the $xy$
plane, which is known as a {\it Penning trap}
\cite{Brown1}. The corresponding Hamiltonian can be written as
\begin{equation}
\Hop = {1\over{2m}} \left[ \pvec-{q\over c}\Avec \right]^{2} +
{{qV_{o}}\over{2d^2}} \left( \zop{}^{2}-{{\xop{}^{2}+\yop{}^{2}}\over2}
\right) \;,
\label{eq:Hcl}
\end{equation}
where $\Avec = (B_{o}y/2,-B_{o}x/2,0)$ is the vector potential,
$c$ is the speed of light, $d$
characterizes the dimensions of the trap, $V_{o}$ is the 
electrostatic potential applied to its electrodes, and $q$ the 
electron charge.
 
The spatial part of the electronic wave function
consists of three degrees of freedom, but neglecting the slow magnetron
motion (whose characteristic frequency lies in the kHz region),
here we only consider the axial
and cyclotron motions, which are two harmonic oscillators
radiating in the MHz and GHz regions, respectively. The spin dynamics
results from the interaction between the magnetic moment of the electron
and the magnetic field, so that the total quantum Hamiltonian is
\begin{equation}
\Hop_{\rm tot} = \hbar\omegaz(\acop_{z}\aop_{z}+1/2)
+ \hbar\omegac(\acop_{c}\aop_{c}+1/2) + 
{{\hbar\omegas}\over2}\sigmaop_{z} \;.
\label{eq:Hzcs}
\end{equation}
In the previous expression we have introduced the lowering
operator for the cyclotron motion
\begin{equation}
\aop_{c} = {1\over2}\left[ {1\over{2\beta\hbar}}
 (\pop_{x} - i\pop_{y}) - i\beta (\xop - i\yop) \right] \;,
\label{eq:opciclo}
\end{equation}
where $\beta=m\omegac/2\hbar$ and $\omegac= qB_{o}/mc$ is the resonance
frequency associated to the cyclotron oscillation. For the axial motion
we have
\begin{equation}
\aop_{z} = \left( {{m\omegaz}\over{2\hbar}} \right)^{1/2}\zop +
i\left( {1\over{2m\hbar\omegaz}} \right)^{1/2}\hat{p}_{z}\;,
\label{eq:opassial}
\end{equation}
where $\omegaz^{2} = qV_{o}/md^{2}$. In the last term of Eq.~(\ref{eq:Hzcs}),
$\sigmaop_{z}$ is the Pauli spin matrix and $\omegas = (g/2)\omegac$.

The obtained Hamiltonian~(\ref{eq:Hzcs}) is then made of three
independent terms. Even though the only physical observable experimentally
detectable is the axial momentum $\pop_{z}$, in the following
both the cyclotron and spin states will be reconstructed. Considering the
eigenstates of $\sigmaop_{z}$
\begin{equation}
|\uparrow \rangle = \left( \begin{array}{c} 1 \\ 0 \end{array} \right)
\;, \qquad
|\downarrow \rangle = \left( \begin{array}{c} 0 \\ 1 \end{array} \right) \;,
\label{eq:suegiu}
\end{equation}
we can write the most general pure state of the trapped electron in the 
form
\begin{equation}
|\Psi\rangle = c_{1}|\psi_{1}\rangle|\uparrow\rangle + 
c_{2}|\psi_{2}\rangle|\downarrow\rangle \;, 
\label{eq:psi1}
\end{equation}
$|\psi_{1}\rangle$ and $|\psi_{2}\rangle$ being two unknown cyclotron
states. The complex coefficients $c_{1}$ and $c_{2}$, satisfying the 
normalization condition $|c_{1}|^{2}+|c_{2}|^{2}=1$, are also to be
determined.

The electronic state (\ref{eq:psi1}) possesses two very interesting
features: first, if the cyclotron states 
$|\psi_{1}\rangle$ and $|\psi_{2}\rangle$ are macroscopically
distinguishable, $|\Psi\rangle$ is a typical example of
{\it Schr\"odinger-cat state}~\cite{zur}. Second, the full state of the
trapped electron is an {\it entangled state} between the spin and cyclotronic
degrees of freedom (unless $|\psi_{1}\rangle=|\psi_{2}\rangle$).
Introducing the total density operator $\hat{R}=|\Psi\rangle\langle\Psi |$
associated to the pure state $|\Psi\rangle$, we can express the 
corresponding total density matrix $R$ in the basis of the 
eigenstates~(\ref{eq:suegiu}) of $\hat{\sigma}_{z}$ in the form
\begin{equation}
R = \left( \begin{array}{cc} 
|c_{1}|^{2}|\psi_{1}\rangle\langle\psi_{1}| &
c_{1}c_{2}^{*}|\psi_{1}\rangle\langle\psi_{2}| \\  &  \\
c_{2}c_{1}^{*}|\psi_{2}\rangle\langle\psi_{1}| &
|c_{2}|^{2}|\psi_{2}\rangle\langle\psi_{2}| \end{array} \right) \;,
\label{eq:Rmat}
\end{equation}
whose elements are operators. Its diagonal elements represent the 
possible cyclotron states, while the off-diagonal ones are the 
quantum coherences and contain information about the quantum 
interference effects due to the entanglement between the spin and 
cyclotron degrees of freedom.

It is also possible to give a phase-space description of the complete 
quantum state of the trapped electron by introducing the Wigner-function 
matrix~\cite{wal} whose elements are given by
\begin{equation}
W_{ij}(\alpha) = \langle \hat{\delta}_{ij} (\alpha -\hat{a})\rangle
={\rm Tr}[\hat{R}\hat{\delta}_{ij}(\alpha -\hat{a})]\;,
\label{eq:wij}
\end{equation}
where $i,j=1,2$ and $\hat{\delta}_{ij}(\alpha -\hat{a})$ is an operator
in the product Hilbert space ${\cal H} = {\cal H}_{\rm cyc} \otimes
{\cal H}_{\rm spin}$ defined as
\begin{equation}
\hat{\delta}_{ij}(\alpha -\hat{a}) = |i\rangle\langle j| 
\hat{\delta} (\alpha - \hat{a})\;.
\label{eq:dij}
\end{equation}
In the previous expression the operator-valued delta function
$\hat{\delta} (\alpha - \hat{a})$ is the Fourier transform of the 
displacement operator $\hat{D}(\alpha)=\exp(\alpha \hat{a}^{\dagger} -
\alpha^{*}\hat{a})$.

\section{Measurement Scheme and Reconstruction Procedure}
\label{measure}

The basic idea of our reconstruction procedure is very simple: Adding 
a particular inhomogeneous magnetic field---known as the ``magnetic bottle''
field~\cite{Brown1}---to that already present in the trap, it is 
possible to perform a simultaneous measurement of both the spin and 
the cyclotronic excitation numbers. Repeated measurements of this 
type allow us to recover the probability amplitudes associated to the two 
possible spin states and the cyclotron probability distribution 
$P(n_{c})$ in the Fock basis. The reconstruction of the cyclotron 
density matrices $\rho_{ii}=|\psi_{i}\rangle\langle\psi_{i}|$ ($i=1,2$)
in the Fock basis is then possible by employing a technique similar to 
the Photon Number Tomography (PNT)~\cite{sman,man} which exploits a 
phase-sensitive reference field that displaces in the phase space the
particular state one wants to reconstruct~\cite{mafo}.

In close analogy with the procedure described in 
Refs.~\cite{Brown1,man}, the coupling between the different degrees 
of freedom in Eq.~(\ref{eq:Hzcs}) is obtained modifying the vector 
potential with the addition of the magnetic bottle field~\cite{Brown1} 
so that $\Avec$ takes the form
\begin{equation}
\Avec = 
\frac{1}{2}\left[-B_{0}\hat{y}-b\left(\hat{y}\hat{z}^{2}
-\frac{\hat{y}^{3}}{3}\right), B_{0}\hat{x} + b\left(\hat{x}\hat{z}^{2}
-\frac{\hat{x}^{3}}{3}\right), 0\right]\;.
\label{eq:avec}
\end{equation}
Such a vector potential gives rise to an interaction term in the total 
Hamiltonian,
\begin{equation}
\hat{H}_{\rm int} = \hbar \kappa \left[\left(\hat{a}^{\dagger}_{c}
\hat{a}_{c} + \frac{1}{2}\right) + 
\frac{g}{2}\frac{\hat{\sigma}_{z}}{2} \right] \hat{z}^{2}\;,
\label{eq:hint}
\end{equation}
where the coupling constant $\kappa=qb/mc$ is directly related to the 
strength $b$ of the magnetic bottle field.

Eq.~(\ref{eq:hint}) describes the the fact that the axial angular 
frequency is affected both by the number of cyclotron excitations 
$\hat{n}_{c}=\hat{a}^{\dagger}_{c}\hat{a}_{c}$ and by the eigenvalue 
of $\hat{\sigma}_{z}$. In terms of the lowering operator $\hat{a}_{z}$ 
for the axial degree of freedom, the Hamiltonian that describes the 
interaction among the axial, cyclotron, and spin motions can be 
written as
\begin{equation}
\hat{H}_{\rm czs} = \hbar \omega_{z}\left(\hat{a}^{\dagger}_{z}\hat{a}_{z}
+ \frac{1}{2}\right)\frac{\hat{\Omega}^{2}_{z}}{\omega^{2}_{z}}\;,
\label{eq:hczsz}
\end{equation}
where the operator frequency $\hat{\Omega}_{z}$ is given by
\begin{equation}
\hat{\Omega}^{2}_{z} = \omega^{2}_{z} + \frac{\hbar \kappa}{m}
\left[\left(\hat{a}^{\dagger}_{c}\hat{a}_{c} + \frac{1}{2}\right) + 
\frac{g}{4}\hat{\sigma}_{z} \right]\;.
\label{eq:Omegaz}
\end{equation}
The operator $\hat{\Omega}_{z}$ is the modified axial frequency which 
can be experimentally measured~\cite{Brown1} after the application of 
the inhomogeneous magnetic bottle field. What is actually measured is 
an electric current (which is proportional to $\hat{p}_{z}$) that 
gives the axial frequency shift $\hat{\Omega}_{z}^{2}$~\cite{Brown1}.
One immediately sees that the 
spectrum of $\hat{\Omega}_{z}$ is discrete: Since the electron $g$
factor is slightly (but measurably~\cite{Brown1}) different from 2,
$\hat{\Omega}_{z}$ assumes a different value for every pair of
eigenvalues of $\hat{n}_{c}$ and $\hat{\sigma}_{z}$.

\section{Spin Measurements}
\label{spin}

If one can perform a large set of measurements of $\hat{\Omega}_{z}$ 
in such a way that before each measurement the state $|\Psi\rangle$ is 
always prepared in the same way, it is possible to recover the 
probabilities $P(\uparrow)$ and $P(\downarrow)$ associated to the two 
possible eigenvalues of $\hat{\sigma}_{z}$, namely $|c_{1}|$ and 
$|c_{2}|$. Recalling Eq.~(\ref{eq:psi1}), we have
\begin{mathletters}
\label{eq:pupdown}
\begin{eqnarray}
P(\uparrow) & = & {\rm Tr}_{\rm cyc}[\langle \uparrow | \Psi\rangle\langle \Psi
|\uparrow\rangle] = |c_{1}|^{2}\;,
\label{eq:pup} \\
P(\downarrow) & = & {\rm Tr}_{\rm cyc}[\langle \downarrow | \Psi\rangle\langle \Psi
|\downarrow\rangle] = |c_{2}|^{2}=1-P(\uparrow)\;.
\label{eq:pdown}
\end{eqnarray}
\end{mathletters}
However, this kind of measurement does not allow to retrieve the 
relative phase $\theta$ between the complex coefficients $c_{1}$ and 
$c_{2}$ in the superposition~(\ref{eq:psi1}). We can then add a 
time-dependent magnetic field $\bvec_{0}(t)$ oscillating in the $xy$ 
plane perpendicular to the trap axis~\cite{Brown1}, {\it i.e.}
\begin{equation}
\bvec_{0}(t)=b_{0}\cos(\omega t) {\bf \hat{x}}
+ b_{0}\sin(\omega t) {\bf \hat{y}}\;.
\label{eq:bzero}
\end{equation}
The resulting interaction Hamiltonian in the interaction picture is
\begin{eqnarray}
\hat{H}_{\rm rot}^{\rm I}(t) & = & \exp\left[\frac{i}{\hbar} 
\hat{H}_{0}t\right] \hat{H}_{\rm rot}(t)\exp\left[-\frac{i}{\hbar} 
\hat{H}_{0}t\right]
\nonumber \\
 & = & \frac{\hbar}{2}[(\omega_{s}-\omega)\hat{\sigma}_{z}
 +\omega_{R}\hat{\sigma}_{x}]\;,
\label{eq:hrotint}
\end{eqnarray}
where
\begin{mathletters}
\begin{eqnarray}
\hat{H}_{0} & = & \frac{\hbar\omega}{2}\hat{\sigma}_{z}\;,
\label{eq:hzero} \\
\hat{H}_{\rm rot}(t) & = & \frac{\hbar}{2}[(\omega_{s}-\omega)\hat{\sigma}_{z}
\nonumber \\
 & & +\omega_{R}(\hat{\sigma}_{x}\cos(\omega t) + 
\hat{\sigma}_{y}\sin(\omega t))]\;.
\label{eq:hrott}
\end{eqnarray}
\end{mathletters}
In the above equations, $\hat{H}_{\rm rot}(t)$ is the interaction 
Hamiltonian in a frame rotating at the driving frequency $\omega$, 
while $\omega_{R}=gqb_{0}/2mc$ is the Rabi frequency.

The evolution of the state~(\ref{eq:psi1}) subjected to the 
Hamiltonian~(\ref{eq:hrotint}) in the resonant case 
$\omega=\omega_{s}$, yields the state
\begin{eqnarray}
|\Psi(\bar{t})\rangle & = & \exp\left[-\frac{i}{\hbar} \hat{H}_{\rm 
rot}^{\rm I} (t) \bar{t}\right] |\Psi\rangle
\nonumber \\
 & = & \frac{\protect\sqrt{2}}{2}
 [(c_{1}|\psi_{1}\rangle - ic_{2}|\psi_{2}\rangle)|\uparrow\rangle
\nonumber \\
 & & + (-ic_{1}|\psi_{1}\rangle +c_{2}|\psi_{2}\rangle)|\downarrow\rangle]\;,
\label{eq:psibar}
\end{eqnarray}
obtained applying the driving field~(\ref{eq:bzero}) for a time
$\bar{t}=\pi/2\omega_{R}$.
We can now repeat the spin measurements just as we have described 
above in the case of the {\em unknown} initial state $|\Psi\rangle$: Soon 
after $\bvec_{0}(t)$ is switched off, the magnetic 
bottle field is applied again and the spin measurement is performed. 
Repeating this procedure over and over again (with the same unknown 
initial state) for a large number of times, it is possible to recover 
the probabilities $\bar{P}(\uparrow)$ and $\bar{P}(\downarrow)$ 
associated to the two spin eigenvalues for the state 
$|\Psi(\bar{t})\rangle$ of Eq.~(\ref{eq:psibar}). Without loss of 
generality, we can assume $c_{1}\in R$, $c_{2}=|c_{2}| e^{i\theta}$, 
and $\langle\psi_{1}|\psi_{2}\rangle = re^{i\beta}$, which yield
\begin{mathletters}
\label{eq:pbar}
\begin{eqnarray}
\bar{P}(\uparrow) & = & \frac{1}{2}[1+2r|c_{1}| |c_{2}| 
\sin(\theta+\beta)]
\label{eq:pbarup} \\
\bar{P}(\downarrow) & = & \frac{1}{2}[1-2r|c_{1}| |c_{2}| 
\sin(\theta+\beta)]\;.
\label{eq:pbardown}
\end{eqnarray}
\end{mathletters}
It is important to note that the probabilities $\bar{P}(\uparrow)$ and 
$\bar{P}(\downarrow)$ can be experimentally sampled and that the 
modulus $r$ and the phase $\beta$ of the scalar product 
$\langle\psi_{1}|\psi_{2}\rangle$ can be both derived from the 
reconstruction of the cyclotron density matrices $\rho_{11}$ and 
$\rho_{22}$, as we shall explain in the next section. Thus we are 
able to find the relative phase $\theta$ by simply inverting one of 
the two Eqs.~(\ref{eq:pbar}), {\it e.g.}
\begin{equation}
\theta = \arcsin \left[\frac{2\bar{P}(\uparrow)-1}{2r |c_{1}| |c_{2}|}
\right]- \beta\;.
\label{eq:theta}
\end{equation}
The resulting $\pi$ ambiguity in the $\arcsin$ function in the right hand
side of Eq.~(\ref{eq:theta}) can be eliminated by choosing a second 
interaction time $\bar{t}'$ and repeating the procedure above.

\section{Tomographic Reconstruction of the Cyclotron States}
\label{tomo}

Let us consider again the state of Eq.~(\ref{eq:psi1}): every time 
$\hat{\Omega}_{z}$ (and therefore $\hat{\sigma}_{z}$) is measured, the 
total wave function is projected onto $|\uparrow\rangle|n_{c}\rangle$ or
$|\downarrow\rangle|n_{c}\rangle$, where $|n_{c}\rangle$ is a cyclotron
Fock state.
We then propose a tomographic reconstruction technique in which the 
state to be measured is combined with a reference field whose complex 
amplitude is externally varied (as it is usually done in optical 
homodyne tomography~\cite{vori,dar,beck,natu}) in order to displace
the {\em 
unknown} density operator in the phase space (a technique very close 
to the PNT scheme~\cite{wal,man,ban}). In particular, we shall sample 
the cyclotron density matrix in the Fock basis by varying only the 
phase $\varphi$ of the reference field, leaving unaltered its 
modulus $|\alpha|$~\cite{sman,mafo,opa}.

Following Ref.~\cite{sman}, immediately before the measurement of 
$\hat{\Omega}_{z}$, we apply to the trap electrodes a driving field 
generated by the vector potential
\begin{equation}
\Avec = \left( 2\frac{mc}{\beta|q|} Im[\epsilon e^{-i\omega_{c}t}],
2\frac{mc}{\beta|q|} Re[\epsilon e^{-i\omega_{c}t}],0\right)\;,
\label{eq:vecpot}
\end{equation}
where $\epsilon$ is the field amplitude, which gives rise to a 
Hamiltonian term of the form
\begin{equation}
\hat{H}_{\rm drive} = -i\hbar (\epsilon e^{-i\omega_{c}t} 
\hat{a}^{\dagger}_{c} - \epsilon^{*} e^{i\omega_{c}t} 
\hat{a}_{c})\;.
\label{eq:hdrive}
\end{equation}
The time evolution of the projected density operator $\hat{\rho}_{ii}$
($i=1,2$) according to the Hamiltonian~(\ref{eq:hdrive}) may then be 
written in the cyclotron interaction picture as
\begin{eqnarray}
\hat{\rho}_{ii}(\epsilon, t) & = & 
\exp\left(\frac{i}{\hbar}\tilde{H}_{\rm drive}t\right)
\hat{\rho}_{ii}(0) \exp\left(-\frac{i}{\hbar}\tilde{H}_{\rm drive}t\right)
\nonumber \\
 & = & \hat{D}^{\dagger}(\alpha) \hat{\rho}_{ii}(0) \hat{D}(\alpha)\;,
\label{eq:rhodisp}
\end{eqnarray}
where we have defined the complex parameter $\alpha=\epsilon t$ ($t$ 
being the interaction time) and $\tilde{H}_{\rm drive}$ is given by
\begin{equation}
\tilde{H}_{\rm drive} = -i\hbar(\epsilon \hat{a}^{\dagger}_{c}
-\epsilon^{*}\hat{a}_{c})\;.
\label{eq:htilde}
\end{equation}
The right-hand side of Eq.~(\ref{eq:rhodisp}) is then the desired displaced 
density operator, where the displacement parameter $\alpha$ is a 
function of both the strength $\epsilon$ of the driving field and the 
interaction time $t$. Thus we can interpret the quantity
\begin{eqnarray}
P^{(i)}(n_{c},\alpha) & = & {\rm Tr}[\hat{\rho}_{ii}(\alpha) 
|n_{c}\rangle\langle n_{c}|]
\nonumber \\
 & = & \langle n_{c}| \hat{D}^{\dagger}(\alpha) \hat{\rho}_{ii}(0)
 \hat{D}(\alpha) |n_{c}\rangle
\nonumber \\
 & = & \langle n_{c},\alpha | \hat{\rho}_{ii}(0) |n_{c},
 \alpha\rangle\;,
\label{eq:pnc}
\end{eqnarray}
as the probability of finding the cyclotron state $|\psi_{i}\rangle$ 
with an excitation number $n_{c}$ after the application of the 
driving field of amplitude $\epsilon$ for a time $t$. Fixing a 
particular value of $\alpha$, and measuring $\hat{n}_{c}$, it is then
possible to recover the probability distribution~(\ref{eq:pnc}) 
performing many identical experiments.

Expanding the density operator $\hat{\rho}_{ii}$ in the Fock basis, 
and defining $N_{c}$ as an appropriate estimate of the maximum number 
of cyclotronic excitations (cut-off), we have
\begin{equation}
P^{(i)}(n_{c},\alpha) = \sum_{k,m=0}^{N_{c}} \langle n_{c}, \alpha 
|k\rangle\langle k|\hat{\rho}_{ii}|m\rangle\langle 
m|n_{c},\alpha\rangle\;.
\label{eq:pinc}
\end{equation}
The projection of the displaced number state $|n_{c},\alpha\rangle$ 
onto the Fock state $|m\rangle$ can be obtained (generalizing the 
result derived in Ref.~\cite{cah}) as
\begin{eqnarray}
\langle m |n,\alpha\rangle & = & \langle m | \hat{D} (\alpha)
|n\rangle
= e^{-|\alpha|^{2}}\sqrt{\frac{\nu !}{\mu !}} |\alpha|^{\mu-\nu}
\nonumber \\
 &  & \times \exp\left\{i(m-n)\left[\varphi-\pi\theta(n-m)\right]\right\}
\nonumber \\
 & & \times  L^{\mu - \nu}_{\nu}(|\alpha|^{2})\;,
\label{eq:mna}
\end{eqnarray}
where $\theta (x)$ is the Heaviside function, $L_{\nu}^{\mu}$ is the 
associated Laguerre polynomial, and $\mu=\max\{m,n\}$, 
$\nu=\min\{m,n\}$. Inserting Eq.~(\ref{eq:mna}) into 
Eq.~(\ref{eq:pinc}), we get
\begin{eqnarray}
P^{(i)}(n,\alpha) & = & e^{-|\alpha|^{2}} \sum_{k,m=0}^{N_{c}}
\sqrt{\frac{\nu ! \bar{\nu}!}{\mu !\bar{\mu}!}}
|\alpha|^{\mu+\bar{\mu}-\nu-\bar{\nu}}
 L^{\mu - \nu}_{\nu}(|\alpha|^{2})
\nonumber \\
 & & \times  e^{i(m-k)\varphi
 -\pi\left[(m-n)\theta(n-m) - (k-n)\theta(n-k)\right]}
\nonumber \\ 
 & & \times L^{\bar{\mu} - \bar{\nu}}_{\bar{\nu}}(|\alpha|^{2})
 \langle k |\hat{\rho}_{ii} | m\rangle\;,
\label{eq:pina}
\end{eqnarray}
where $\bar{\mu}=\max\{k,n\}$ and $\bar{\nu}=\min\{k,n\}$.

Let us now consider, for a given value of $|\alpha|$, $P^{(i)}(n,\alpha)$
as a function of $\varphi$ and calculate the coefficients of the 
Fourier expansion
\begin{equation}
P^{(s,i)} (n,|\alpha|) = \frac{1}{2\pi}\int_{0}^{2\pi} d\varphi \;
P^{(i)}(n,\alpha) e^{i\varphi}\;,
\label{eq:psina}
\end{equation}
for $s=0,1,2,\ldots$. Combining Eqs.~(\ref{eq:pina}) and 
(\ref{eq:psina}), we get
\begin{equation}
P^{(s,i)} (n,|\alpha|) = \sum_{m=0}^{N_{c}-s} {G}^{(s)}_{n,m}
(|\alpha|) \langle m+s | \hat{\rho}_{ii} | m \rangle\;,
\label{eq:psinag}
\end{equation}
where we have introduced the matrices
\begin{eqnarray}
{G}^{(s)}_{n,m} (|\alpha|) & = & e^{i\pi[(s+m-n)\theta(n-s-m)
- (m-n)\theta(n-m)]}
\nonumber \\
 & \times & e^{-|\alpha|^{2}}
\left(\frac{\nu ! \bar{\nu} !}{\mu ! \bar{\mu} !}\right)^{\frac{1}{2}}
|\alpha|^{\mu+\bar{\mu}-\nu-\bar{\nu}}
 L^{\mu - \nu}_{\nu}(|\alpha|^{2})
\nonumber \\
& \times & L^{\bar{\mu} - \bar{\nu}}_{\bar{\nu}}(|\alpha|^{2})
\;,
\label{eq:gina}
\end{eqnarray}
with $\bar{\mu}=\max\{m+s,n\}$ and $\bar{\nu}=\min\{m+s,n\}$.

We may now note that if the distribution $P^{(i)}(n,\alpha)$ is 
measured for $n\in [0,N]$ with $N\geq N_{c}$, then 
Eq.~(\ref{eq:psinag}) represents for each value of $s$ a system of 
$N+1$ linear equations between the $N+1$ measured quantities and the 
$N_{c}+1-s$ unknown density matrix elements. Therefore, in order to 
obtain the latter, we only need to invert the system
\begin{equation}
\langle m+s | \hat{\rho}_{ii} | m \rangle = \sum_{n=0}^{N}
M^{(s)}_{m,n} (|\alpha|) P^{(s,i)} (n,|\alpha|)\;,
\label{eq:msrhom}
\end{equation}
where the matrices $M$ are given by $M=(G^{T} G)^{-1} G^{T}$.
It is possible to see that these matrices satisfy the relation
\begin{equation}
\sum_{n=0}^{N} M_{m',n}^{(s)} (|\alpha|) G^{(s)}_{n,m} (|\alpha|)
= \delta_{m',m}\;,
\label{eq:complete}
\end{equation}
for $m,m'=0,1,\ldots,N_{c}-s,\ldots$, which means that from the exact 
probabilities satisfying Eq.~(\ref{eq:psinag}) the correct density 
matrix elements are obtained. Furthermore, combining 
Eqs.~(\ref{eq:psina}) and (\ref{eq:msrhom}), we find
\begin{equation}
\langle m+s | \hat{\rho}_{ii} | m \rangle = \frac{1}{2}
\sum_{n=0}^{N} \int_{0}^{2\pi} d\varphi \; M^{(s)}_{m,n} (|\alpha|)
e^{is\varphi} P^{(i)} (n,\alpha)\;,
\label{eq:msrhim}
\end{equation}
which may be regarded as the formula for the direct sampling of the 
cyclotron density matrix. In particular, Eq.~(\ref{eq:msrhim}) clearly 
shows that the determination of the cyclotron state requires only the 
value of $\varphi$ to be varied.

We now only need to reconstruct the off-diagonal parts of the total 
density matrix~(\ref{eq:Rmat}), {\it i.e.} 
$\hat{\rho}_{12}=\hat{\rho}_{21}^{\dagger}
=|\psi_{1}\rangle\langle \psi_{2}|$. This can easily be done under the 
assumption that the initial {\em unknown} electron state is pure, as 
in Eq.~(\ref{eq:psi1}). Then, we have
\begin{equation}
(\rho_{12})_{n,m}=
\frac{\sum_{i=0}^{N}(\rho_{11})_{n,i}(\rho_{22})_{i,m}}{\langle\psi_{1}
|\psi_{2}\rangle}\;.
\label{eq:rho12}
\end{equation}
Writing $|\psi_{1}\rangle = \sum_{i=0}^{N} a_{n} |n\rangle$ and
$|\psi_{2}\rangle = \sum_{i=0}^{N} b_{n} |n\rangle$ in the Fock basis,
we can obtain the desired coefficients from the recursive relation
\begin{equation}
a_{n}=\left[\frac{(\rho_{11})_{n-1,n}}{a_{n-1}}\right]^{*}\;,
\label{eq:an}
\end{equation}
where without loss of generality we can set 
$a_{1}=[(\rho_{11})_{1,1}]^{1/2} \in R$.  A similar relation yields 
the coefficients $b_{n}$ of $|\psi_{2}\rangle$. Finally, the scalar 
product in Eq.~(\ref{eq:rho12}) may be determined as
\begin{equation}
\langle\psi_{1} |\psi_{2}\rangle\ = \sum_{i=1}^{N} a_{i}^{*} b_{i}\;.
\label{eq:scal}
\end{equation}

\section{Simulations and Results}
\label{simu}

In this section we show the results of numerical Monte-Carlo 
simulations of the method presented above, which allow us to state 
that this technique may be quite accurate also in the experimental 
implementation. To account for actual experimental conditions, we 
have considered the effects of a non-unit quantum efficiency $\eta$ in 
the counting of cyclotronic excitations. When $\eta < 1$, the actually 
measured distribution ${\cal P} (k,\alpha)$ is related to the ideal 
distribution $P(n,\alpha)$ by the binomial convolution~\cite{scula}
\begin{equation}
{\cal P}(k,\alpha) = \sum_{n=k}^{\infty} {\cal B}_{k,n} (\eta) 
P(n,\alpha)\;,
\label{eq:pika}
\end{equation}
where
\begin{equation}
{\cal B}_{k,n} (\eta) = \left( \begin{array}{c} n \\ k \end{array} \right)
\eta^{k} (1-\eta)^{n-k}\;.
\label{eq:bino}
\end{equation}
Eq.~(\ref{eq:psinag}) is then modified as
\begin{equation}
{\cal P}^{(s,i)} (n, |\alpha|) = \sum_{m=0}^{N_{c}-s} {\cal G}^{(s)}_{n,m}
(|\alpha|, \eta) \langle m+s | \hat{\rho}_{ii} | m\rangle\;,
\label{eq:psinacal}
\end{equation}
where ${\cal P}^{(s,i)}(n,|\alpha|)$ is again defined according to 
Eq.~(\ref{eq:psina}), but now with ${\cal P}^{(i)} (n,\alpha)$ in 
place of $P^{(i)} (n,\alpha)$. In addition, ${\cal G}^{(s)}_{n,m}
(|\alpha|, \eta)$ is defined as
\begin{equation}
{\cal G}^{(s)}_{n,m} (|\alpha|, \eta) = \sum_{k=0}^{\infty}
{\cal B}_{k,n} (\eta) G_{k,m}^{(s)} (|\alpha|)\;.
\label{eq:geta}
\end{equation}
The matrices ${\cal G}^{(s)}_{n,m} (|\alpha|, \eta)$ can then be 
inverted in the way described in the previous section to obtain the 
matrices ${\cal M}^{(s)}_{n,m} (|\alpha|,\eta)$ that can be used to 
reconstruct the density matrix.

In the above reconstruction procedure, we have to consider two possible
sources of errors associated to any actual measurement process. First, 
we have noticed a strong correlation between the statistical error in 
the simulated reconstruction and the absolute value $|\alpha|$ of 
the coherent field amplitude applied to drive the cyclotron motion.
When $|\alpha|$ is small, the density matrix elements near to the 
main diagonal are accurately reconstructed. Progressively 
increasing $|\alpha|$, the reconstruction of the off-diagonal 
elements becomes more accurate, while the elements close to the 
diagonal present significant fluctuations: to compensate for this 
effect, it is necessary to increase the number of measurements as much 
as possible. Another source of error stems from the truncation of the 
reconstructed density matrix at the value $N_{c}$ in the Fock space.
Neglecting the terms with $n>N_{c}-s$ in Eq.~(\ref{eq:psinag}) causes 
a systematic error. This error can be reduced by increasing $N_{c}$, 
which however may also cause an increase of the statistical error.
This suggests that for a given number of measurement events there is 
an optimal value of $N_{c}$ for which the systematic error is reduced 
below the statistical error.

In the following we will present two examples of application of the 
above method. We shall show the simulated tomographic reconstructions 
of an entangled electronic state of the type
\begin{equation}
|\Psi\rangle = c_{1} |\gamma\rangle |\uparrow\rangle
+ c_{2} e^{i\theta} |\gamma e^{i\xi}\rangle | \downarrow\rangle\;,
\label{eq:state}
\end{equation}
in which $c_{1}$, $c_{2}$, $\theta$, $\gamma$, and $\xi$ are real 
parameters, and $|\gamma\rangle$ is a coherent state of the cyclotron 
oscillator. When $\xi\neq 0$, the state~(\ref{eq:state}) is 
entangled, and when $\xi=\pi$ (with $|\gamma|\gg 1$) it represents 
the most ``genuine'' example of a Schr\"odinger-cat state~\cite{schr}.
We shall display the results of our simulations both in terms of the 
density matrix of Eq.~(\ref{eq:Rmat}) and of the Wigner-function
matrix~\cite{wal,wig} of Eq.~(\ref{eq:wij}). The latter can be derived from 
the density matrix elements $(\rho_{ij})_{n,m}$ through the 
relation~\cite{vito}
\begin{eqnarray}
W_{ij}(x,y) & = & \frac{2}{\pi} \sum_{n} (\rho_{ij})_{n,n} (-1)^{n}
e^{-2(x^{2}+y^{2})} L_{n}[4(x^{2}+y^{2})]
\nonumber \\
 & + & \frac{4}{\pi}\sum_{n\neq m} (\rho_{ij})_{n,m} (-1)^{n}
 \sqrt{\frac{n!}{m!}} [2(x+iy)]^{m-n}
\nonumber \\
 & \times & e^{-2(x^{2}+y^{2})}
 L_{n}^{m-n}[4(x^{2}+y^{2})]\;.
\label{eq:wxy}
\end{eqnarray}

\begin{figure}[t]
\centerline{\epsfig{figure=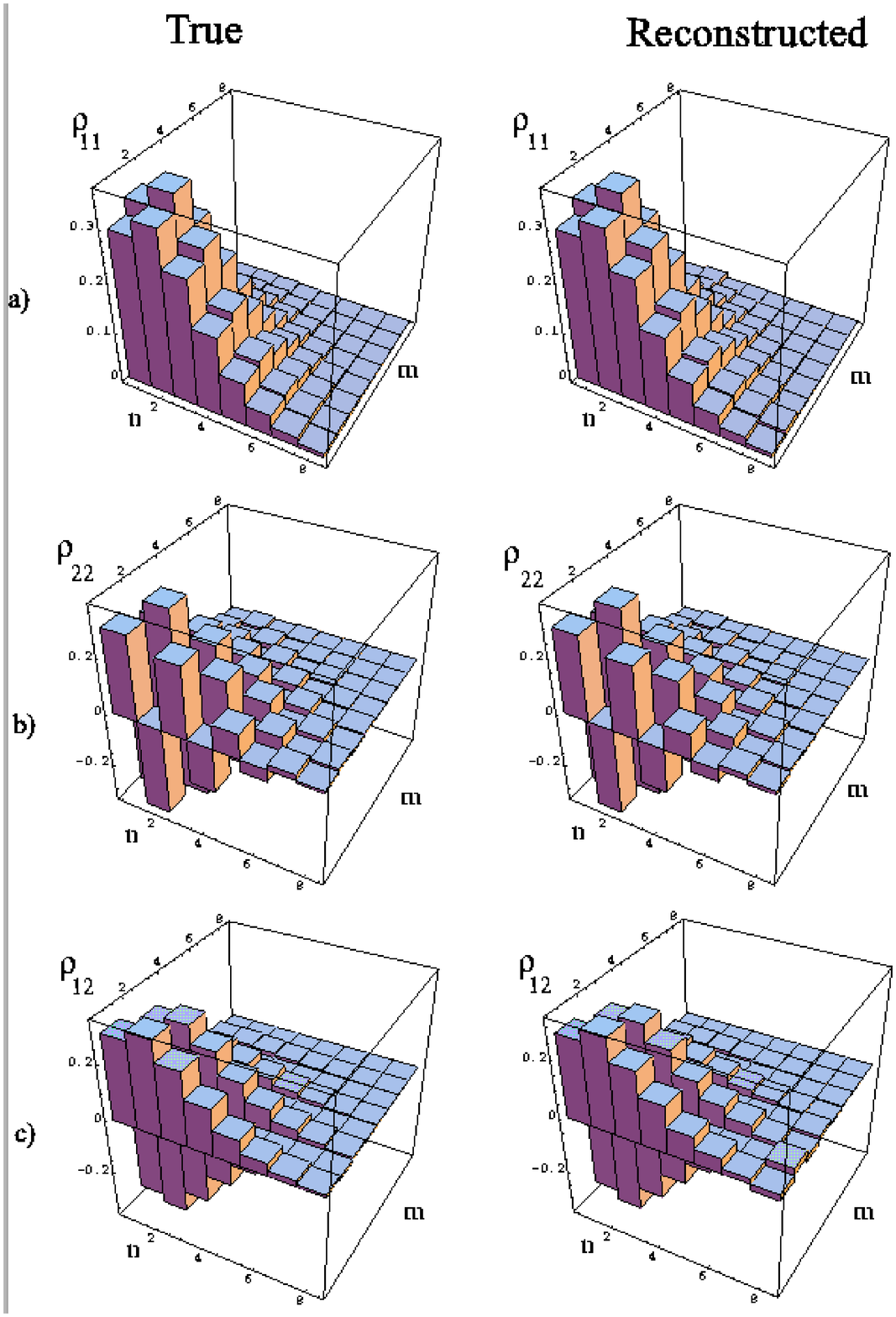,width=14cm}}
\caption{Simulated tomographic reconstruction of the density matrix for the 
state of Eq.~(\protect\ref{eq:state}) with $c_{1}=1/2$, 
$c_{2}=\protect\sqrt{3}/2$, $\theta=\pi$, $\xi=\pi$, and $\gamma=1$. 
The quantum efficiency is $\eta=0.9$ and 10$^{6}$ data per phase have 
been simulated. In this simulation an amplitude $|\alpha|=0.7$ of the 
applied reference field has been used. a) $(\rho_{11})_{n,m}$, b) 
$(\rho_{22})_{n,m}$, and c) $(\rho_{12})_{n,m}$. Here and in the following 
figures the ideal distributions are displayed on the left, whereas 
the reconstructed ones are displayed on the right.}
\label{fg:one}
\end{figure}

\begin{figure}[t]
\centerline{\epsfig{figure=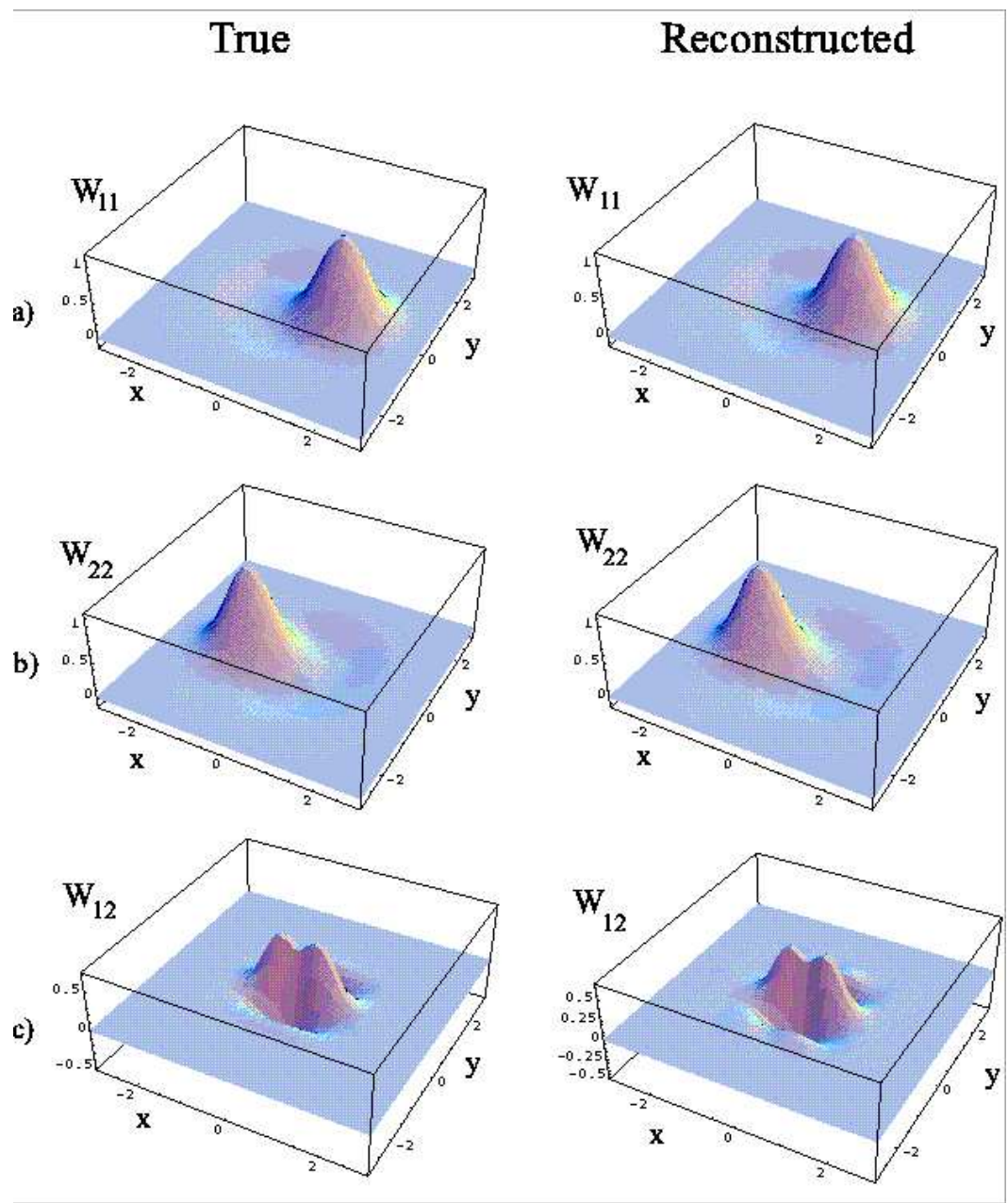,width=14cm}}
\caption{Simulated tomographic reconstruction of the Wigner functions for the 
state of Eq.~(\protect\ref{eq:state}) with the same parameters as in 
Fig.~\protect\ref{fg:one}. 
a) $W_{11}(x,y)$, b) $W_{22}(x,y)$, and c) $W_{12}(x,y)$.}
\label{fg:two}
\end{figure}

In Figs.~\ref{fg:one} and \ref{fg:two} we show the numerical 
reconstruction of an asymmetric superposition of the type of
Eq.~(\ref{eq:state}) with $c_{1}=1/2$, $c_{2}=\sqrt{3}/2$, 
$\theta=\pi$, $\xi=\pi$, $\eta=0.9$ and $\gamma=1$. In Fig.~\ref{fg:one}
we plot the results concerning the density matrix, while in Fig.~\ref{fg:two} 
those concerning the Wigner function. In each figure, the true 
distributions are depicted on the left next to the corresponding 
reconstructed ones. Both the density matrices and the Wigner functions
are well reconstructed.

\begin{figure}[t]
\centerline{\epsfig{figure=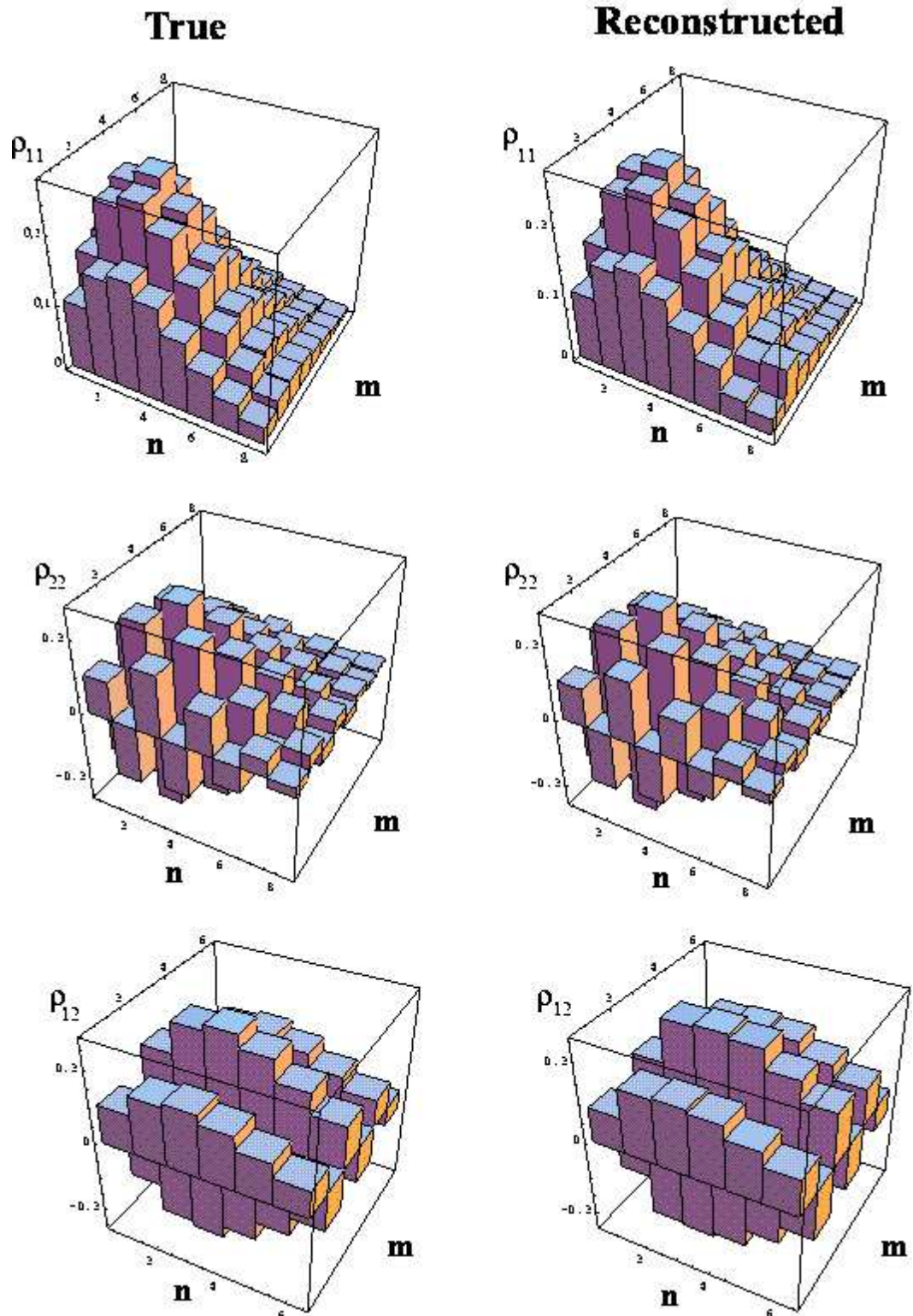,width=14cm}}
\caption{Simulated tomographic reconstruction of the density matrix for the 
state of Eq.~(\protect\ref{eq:state}) with $c_{1}=c_{2}=\protect\sqrt{2}/2$,
$\theta=0$, $\xi=\pi$, and $\gamma=1.5$. 
The quantum efficiency is $\eta=0.9$ and 10$^{6}$ data per phase have 
been simulated. In this simulation an amplitude $|\alpha|=1.2$ of the 
applied reference field has been used. Top: $(\rho_{11})_{n,m}$; Middle:
$(\rho_{22})_{n,m}$; Bottom: $(\rho_{12})_{n,m}$.}
\label{fg:three}
\end{figure}

\begin{figure}[t]
\centerline{\epsfig{figure=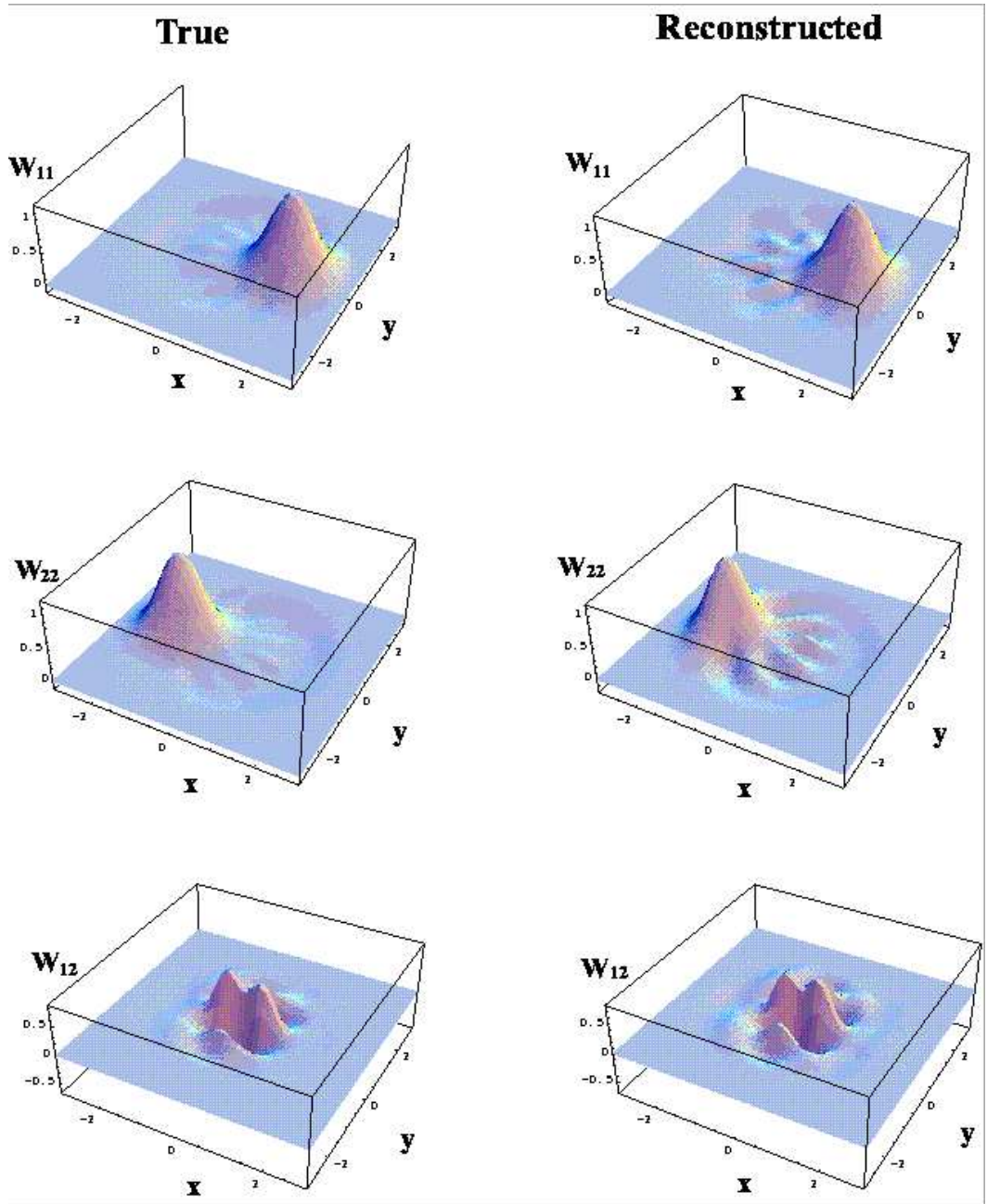,width=14cm}}
\caption{Simulated tomographic reconstruction of the Wigner functions for the 
state of Eq.~(\protect\ref{eq:state}) with the same parameters as in 
Fig.~\protect\ref{fg:three}. 
Top: $W_{11}(x,y)$; Middle: $W_{22}(x,y)$; Bottom: $W_{12}(x,y)$.}
\label{fg:four}
\end{figure}

In Figs.~\ref{fg:three} and \ref{fg:four} analogous results are shown 
for a symmetric superposition state~(\ref{eq:state}) with 
$c_{1}=c_{2}=\sqrt{2}/2$, $\theta=0$, $\xi=\pi$, $\eta=0.9$ and $\gamma=1.5$.
Again, the reconstruction is faithful. We would like to emphasize the 
particular shape of $\rho_{12}$ and $W_{12}$ in both the examples 
above. It is due to the quantum interference given by the entanglement
between the two degrees of freedom: in fact, in absence of 
entanglement ($\xi=0$) $\rho_{12}$ would just be a replica of the 
diagonal parts $\rho_{11}$ and $\rho_{22}$. In both the cases 
considered (Figs.~\ref{fg:one} and \ref{fg:three}) the off-diagonal 
density matrix elements are real, due to the particular choice of the 
cyclotron states. The imaginary parts of the reconstructed density 
matrices turn out to be smaller than 10$^{-3}$.

We have performed a large number of simulations with different states 
and several values of the parameters, which confirm that the present 
method is quite stable and accurate. In addition, and for all the 
cases considered, the values of the parameters $c_{1}$, $c_{2}$, and 
$\theta$ (which obviously do not enter the plots of 
Figs.~\ref{fg:one}--\ref{fg:four}) are very well recovered, with a 
relative error of the order of $10^{-5}$.

\section{Conclusions}
\label{conclu}

In this paper we have proposed a technique suitable to reconstruct 
the (entangled) state of the cyclotron and spin degrees of freedom of 
an electron trapped in a Penning trap. It is based on the magnetic 
bottle configuration, which allows simultaneous measurements of the 
spin component along the $z$ axis and of the cyclotron excitation number.
The cyclotron state is reconstructed with the use of a 
tomographic-like method, in which the phase of a reference driving 
field is varied. The numerical results based on Monte-Carlo 
simulations indicate that even in the case of a non-unit quantum 
efficiency the reconstructed density matrices and Wigner functions are 
almost identical to the ideal distributions. An experimental 
implementation of the proposed method might yield new insight in the 
foundations of quantum mechanics and allow further progress in the 
field of quantum information~\cite{ana}.

\acknowledgments

We gratefully thank G.~M.~D'Ariano for help with the numerical 
simulations. This work has been partially supported by INFM (through 
the 1997 Advanced Research Project ``CAT''), by the
European Union in the framework of the TMR Network ``Microlasers
and Cavity QED'', and by MURST under the ``Cofinanziamento 1997''.

\end{multicols}

\begin{multicols}{2}

\end{multicols}

\widetext

\end{document}